\documentclass[fleqn,10pt]{wlscirep}
\usepackage[utf8]{inputenc}
\usepackage[T1]{fontenc}
\usepackage{orcidlink}
\usepackage{amsthm}
\title{Emergence of power laws in hierarchical dynamics on multi-level graphs}

\author[1,2,*,+]{Tommaso Rondini\,\orcidlink{0009-0001-2867-212X}}
\author[1,2,*,+]{Gregorio Berselli,\orcidlink{0009-0004-7797-9138}}
\author[1,2]{Mirko Degli Esposti,\orcidlink{0000-0003-0316-3449}}
\author[1,2]{Armando Bazzani\orcidlink{0000-0002-9633-0017}}
\affil[1]{University of Bologna, Department of Physics and Astronomy, Bologna, 40126, Italy}
\affil[2]{INFN, Bologna, 40127, Italy}

\affil[*]{\url{tommaso.rondini2@unibo.it} \url{gregorio.berselli2@unibo.it}}

\affil[+]{these authors contributed equally to this work}

\begin{abstract}
Power law distributions are widely recognized in complex systems as indicative of underlying complexity in interaction networks and critical macroscopic behavior. Previous studies have emphasized the importance of network structure and dynamics in understanding the emergence of such statistical patterns and predicting extreme events. In this study, we investigate the emergence of power law behavior in delay distributions within a multi-level hierarchical network of agents governed by priority rules. Using railway systems as case study, we model the dynamics of high-speed and local trains agents assigned distinct priority levels. By introducing stochastic fluctuations into scheduled travel times, derived from empirical data, we observe that local trains exhibit a markedly higher incidence of larger delays than high-speed trains. We propose a queue-based dynamical model, calibrated using Italian railway data, and validate our findings through comparative analysis with Italian and German datasets. The model reproduces the empirically observed power law exponent associated with the Italian local train delays. Furthermore, we analyze the influence of operational policies, such as priority assignment and delay compensation thresholds, finding their effects both in data and in the model. These results underscore the capacity of simple hierarchical structures and rule-based dynamics to generate complex statistical behaviors without intricate interaction networks.
\end{abstract}
\begin{document}

\flushbottom
\maketitle
\thispagestyle{empty}


\section*{Introduction}\label{sec1}

The power law distributions have been considered a fingerprint of complexity in the framework of complex system Physics since they can be related to the presence of critical phenomena and long‐range correlations~\cite{bak2013nature, albert2002statistical}. In the context of interaction networks, such distributions have been linked to scale‐free topologies and self‐organized criticality, as exemplified by the Barabási–Albert model of network growth~\cite{barabasi1999emergence} and sandpile models on lattices~\cite{dhar2006theoretical}. But the power law distributions emerge also when one considers dynamical systems on networks, such as in mobility models on transport networks~\cite{zhao2015,umemoto2018}.
Empirical analyses~\cite{newman2005power, boccaletti2014structure} further suggest that understanding the structure and dynamics of real‐world networks is indispensable for predicting extreme events, ranging from large‐magnitude earthquakes to systemic failures in infrastructural systems.

In a previous paper~\cite{mizzi2023}, we suggest as the hierarchical structure of transport network~\cite{alessandretti2020} can explain the emergence of the power law distributions observed in human mobility.
In this work we cope with the problem of understanding the onset of a power law distribution in the delay time of agents moving into a hierarchical transport network. This problem is interesting for the applications to railway transportation networks, where heterogeneous agents (trains) navigate a shared infrastructure under constraints imposed by scheduling, priority rules, and operational policies.

We first analyzed a minimal hierarchical framework based on max-plus algebra~\cite{GOVERDE, maxp} in which agents are assigned to one of two priority classes, namely high‐priority versus low-priority, and interact solely through queue‐discipline rules at common nodes of the two layers (network bottlenecks). The agent dynamics is stochastic and the travel time fluctuations are modeled via a Laplacian noise simulating that each agent moves trying to maintain a specific speed between the nodes that is subject to random fluctuations. We study the model on a two-layers grid network where agents move with a random choice of the origin-destination.
The simulations show that such a simple model naturally yields power‐law delay distributions for low‐priority agents, without invoking complex network topologies or critical tuning of parameters.
Therefore we conjecture that the same mechanism can be applied to explain the delay distribution of regional and local trains that have a lower priority than the high-speed trains. The statistical features of the train delay distribution have been considered by various authors~\cite{yang2019} and different approaches have been proposed to explain the delay distribution~\cite{corman2018} and to study the delay propagation in the network\cite{meester2007}. 
The understanding of the delay distribution is a key issue to develop control strategies that improve the transport network efficiency.
There are many studies about trains dynamic and delay mechanisms~\cite{GOVERDE, monechi2018complex, dekker} that are interested in delay propagation analyzes and predictions. In our work we are interested in the statistical effects of the policy-rules, not the effects on single trains. Indeed, the railway network is just a case study of how hierarchical dynamics on multi-layer graph evolve.
Empirical investigations have documented non‐Gaussian delay distributions across various national networks, with pronounced heavy tails for commuter and regional services~\cite{buchel2020empirical, briggs2007modelling}. Such findings have motivated the development of stochastic and queuing‐based models aiming to reproduce observed delay statistics~\cite{monechi2018complex, gonzalez2008understanding, lian2023bi}. However, prevailing approaches often assume intricate interaction kernels or detailed timetable dependencies, leaving open the question of whether simple hierarchical rules suffice to generate power‐law behavior.
According to the model simulations we cope with this problem by calibrating the queue‐dynamics model using performance metrics from the Italian railway system and comparing the simulations with the delay data collected from both Italian and German railway networks. We further explore how operational policies, particularly delay‐compensation thresholds linked to passenger‐refund eligibility, imprint characteristic cut‐offs in the tail distributions. Our results highlight that the simple hierarchical interaction and policy‐driven rules can play a central role in shaping complex statistical patterns in infrastructural networks.

The paper is organized as follows: in the second section we outline the main results; in the third section we discuss the model, the validation procedure on real railway networks and the main features of the available data. Finally we discuss the relevance of the results and the possible future developments in the last section.  

\section*{Results}\label{secDiscussion}
\begin{figure}[tbp]
    \centering
    \includegraphics[width=\linewidth]{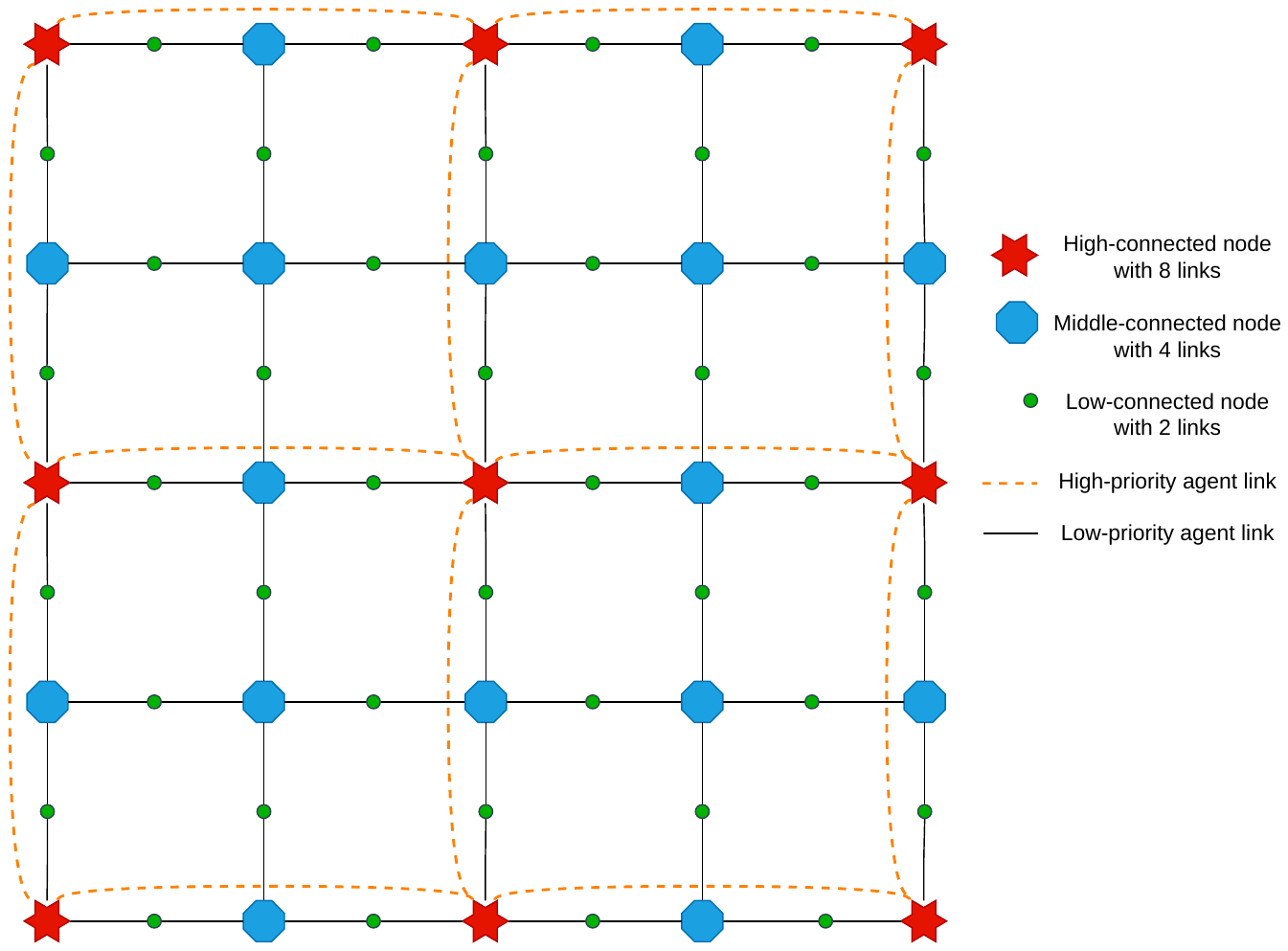}
    \caption{\textbf{Grid network structure.} The figure shows the structure of the grid network with the links between nodes. It is a $3\times 3$ grid example.}
    \label{fig:grid}
\end{figure}

\begin{figure}[tbp]
    \centering
    \includegraphics[width=\linewidth]{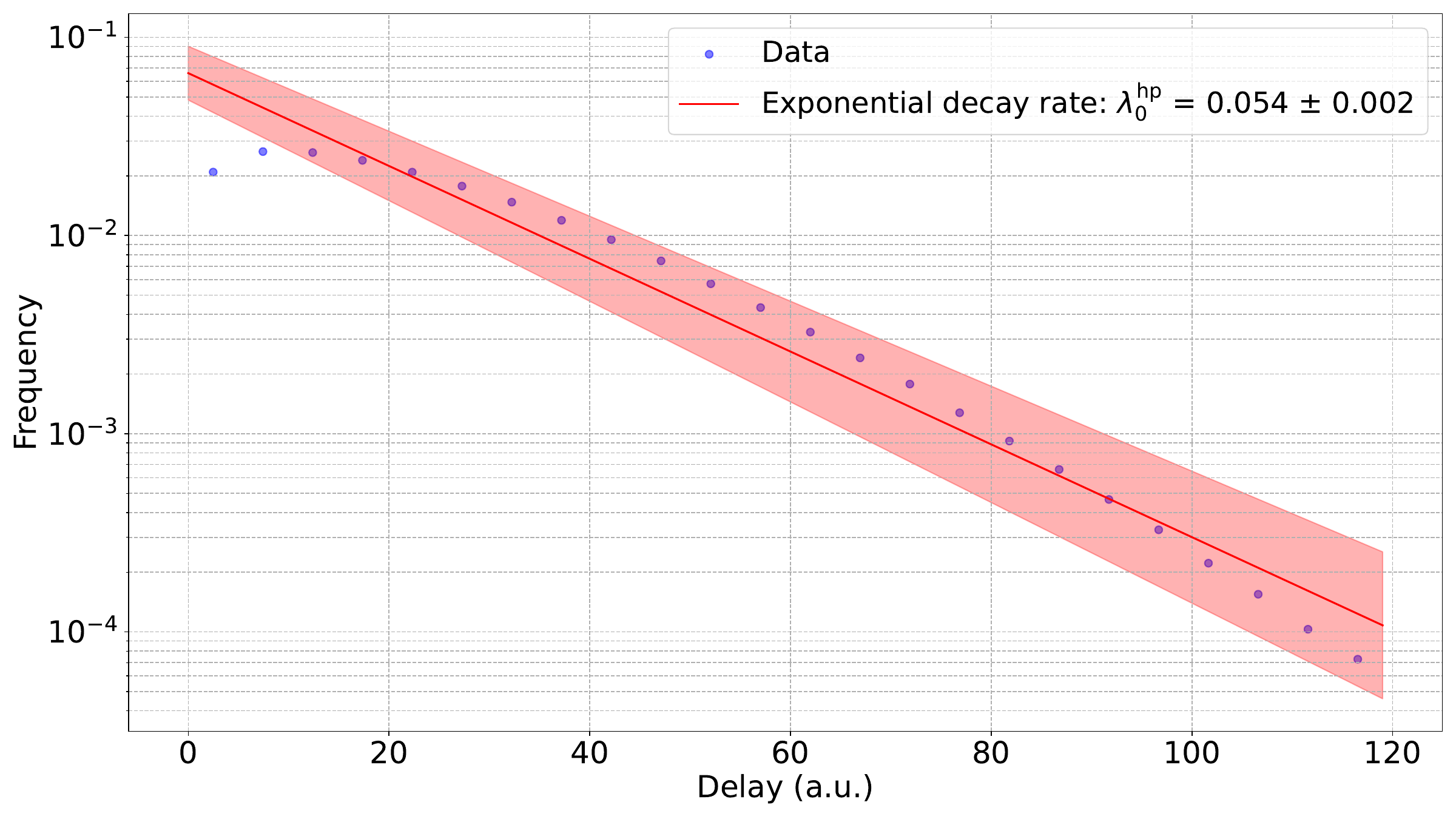}
    \caption{\textbf{Exponential decay of delay distribution for high-priority agents on grid without priority.} The figure shows the distribution of high-priority agent delays without priority, which follows an exponential decay. The red line indicates the exponential fit, calculated and plotted over the same delay range with confidence level of $95\%$.}
    \label{fig:sfr_gr_npr}
\end{figure}

\begin{figure}[tbp]
    \centering
    \includegraphics[width=\linewidth]{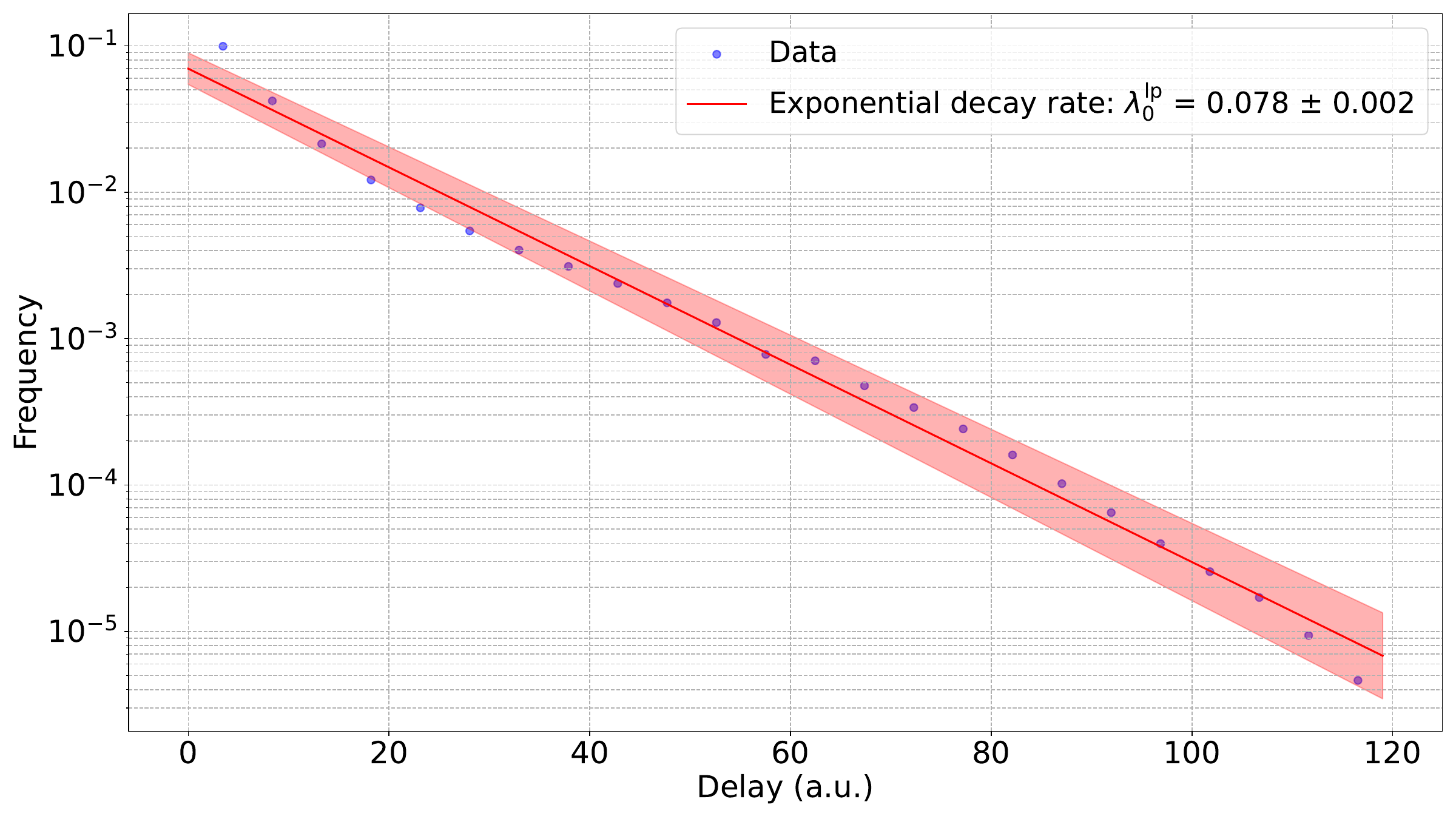}
    \caption{\textbf{Exponential decay of delay distribution for low-priority agents on grid without priority.} The figure shows the distribution of low-priority agent delays without high-priority agent priority, which follows an exponential decay. The red line indicates the exponential fit, calculated and plotted over the same delay range with confidence level of $95\%$.}
    \label{fig:srg_gr_npr}
\end{figure}

\begin{figure}[tbp]
    \centering
    \includegraphics[width=\linewidth]{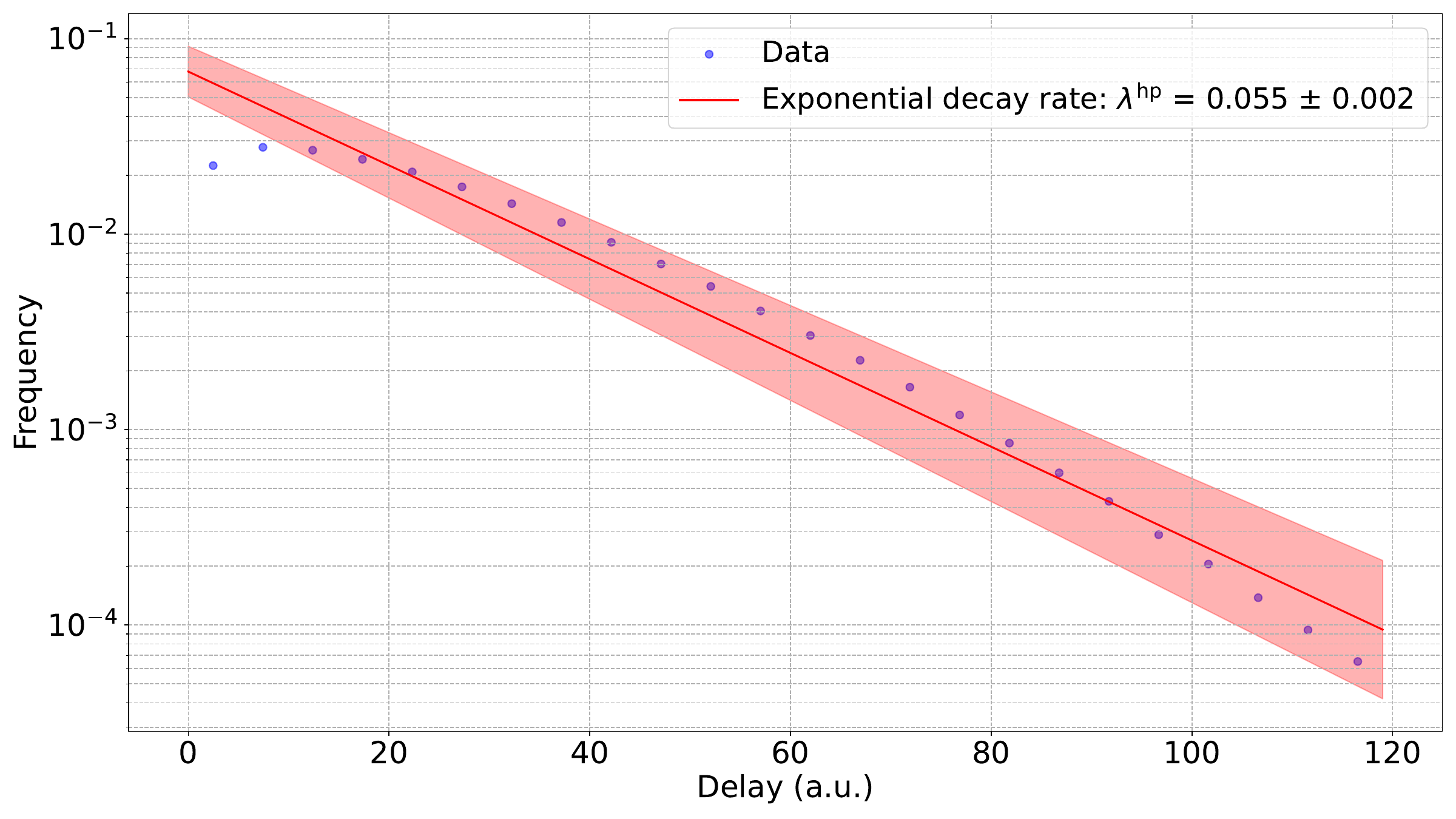}
    \caption{\textbf{Exponential decay of delay distribution for high-priority agents on grid.} The figure shows the distribution of high-priority agent delays, which follows an exponential decay. The red line indicates the exponential fit, calculated and plotted over the same delay range with confidence level of $95\%$.}
    \label{fig:sfr_gr_pr}
\end{figure}

\begin{figure}[tbp]
    \centering
    \includegraphics[width=\linewidth]{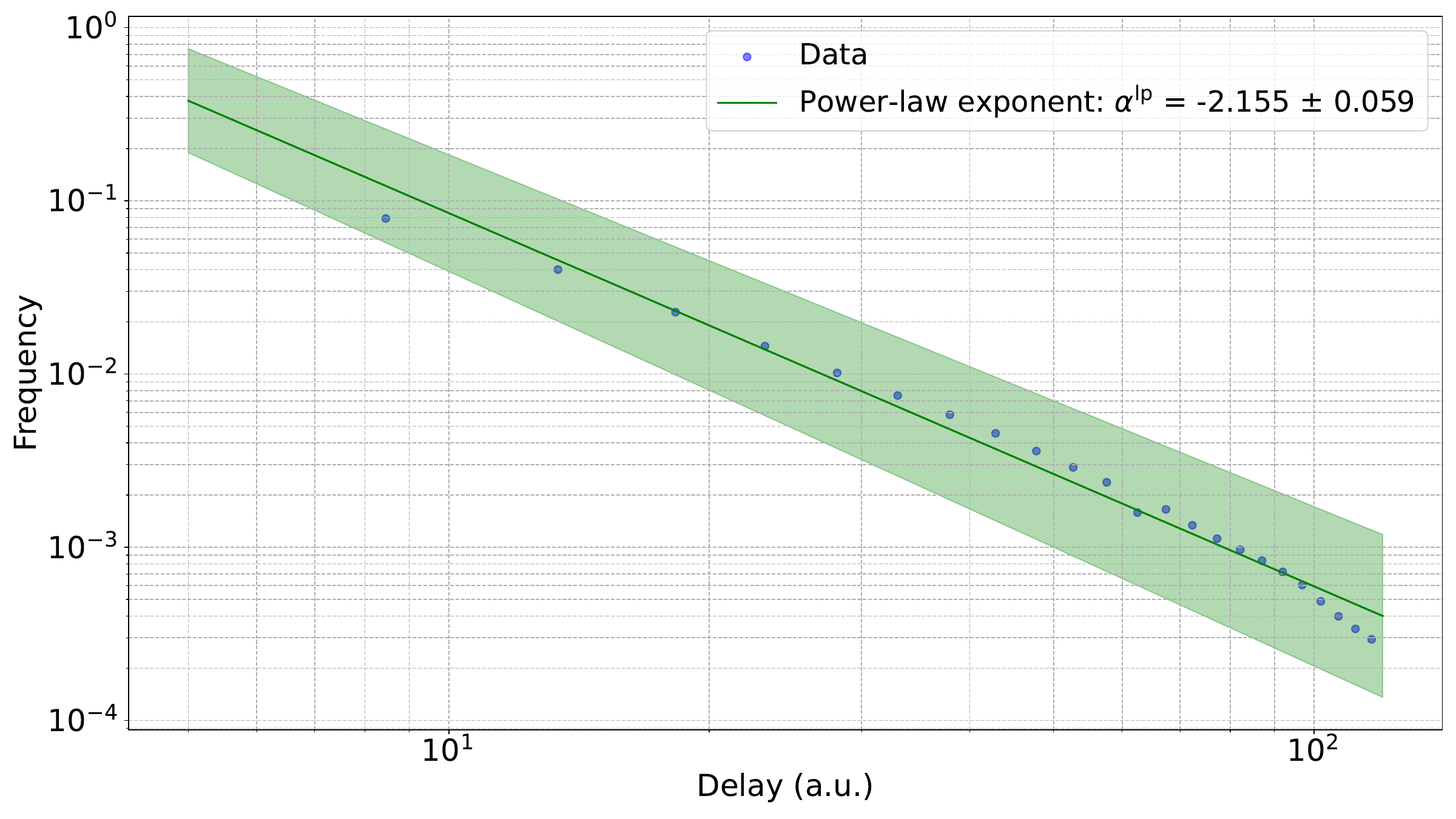}
    \caption{\textbf{Power law decay of delay distribution for low-priority agents on grid.} The figure shows the distribution of low-priority agent delays, which follows a power law decay. The green line indicates the power law fit, calculated and plotted over the same delay range with confidence level of $95\%$.}
    \label{fig:srg_gr_pr}
\end{figure}

To understand the emergence of delays in a transport networks we have initially simulated a grid-like framework composed by two layers (see \autoref{fig:grid}) where the agents perform a random origin-destination dynamics with different priority according to the layer they belong and random delays are introduced using an asymmetrical Laplacian distribution. We consider two scenarios: in the first scenario there is one moving class (i.e., with no priority difference between agents); in the second scenario we give a higher priority to agents in the faster layer. To obtain enough statistics, each simulation has been executed $500$ times with different random seeds.
In the first scenario exponential delay distributions are observed for both types of agents: those that stop only on high-connected nodes with a coefficient $\lambda_0^{hp} = \left(0.054 \pm 0.002\right)$ (\autoref{fig:sfr_gr_npr}) and those that connect all nodes with a coefficient $\lambda_0^{lp} = \left(0.078 \pm 0.002\right )$ (\autoref{fig:srg_gr_npr}).
In the second scenario the agents moving in the fast layer have priority with respect to the agents in the slower layer. The simulations highlight that, whilst the high-priority agent delays still follow an exponential distribution with the same coefficient as in the first scenario ($\lambda^{hp} = \left(0.055 \pm 0.002\right)$), the low-priority agent delays follow a power law distribution with exponent $\alpha^{lp} = \left(-2.155 \pm 0.059\right)$, as reported in \autoref{fig:sfr_gr_pr} and \autoref{fig:srg_gr_pr}.
We conjecture that the exponent depends on the geometrical structure of the transport network.

\begin{figure}[tbp]
    \centering
    \includegraphics[width=\linewidth]{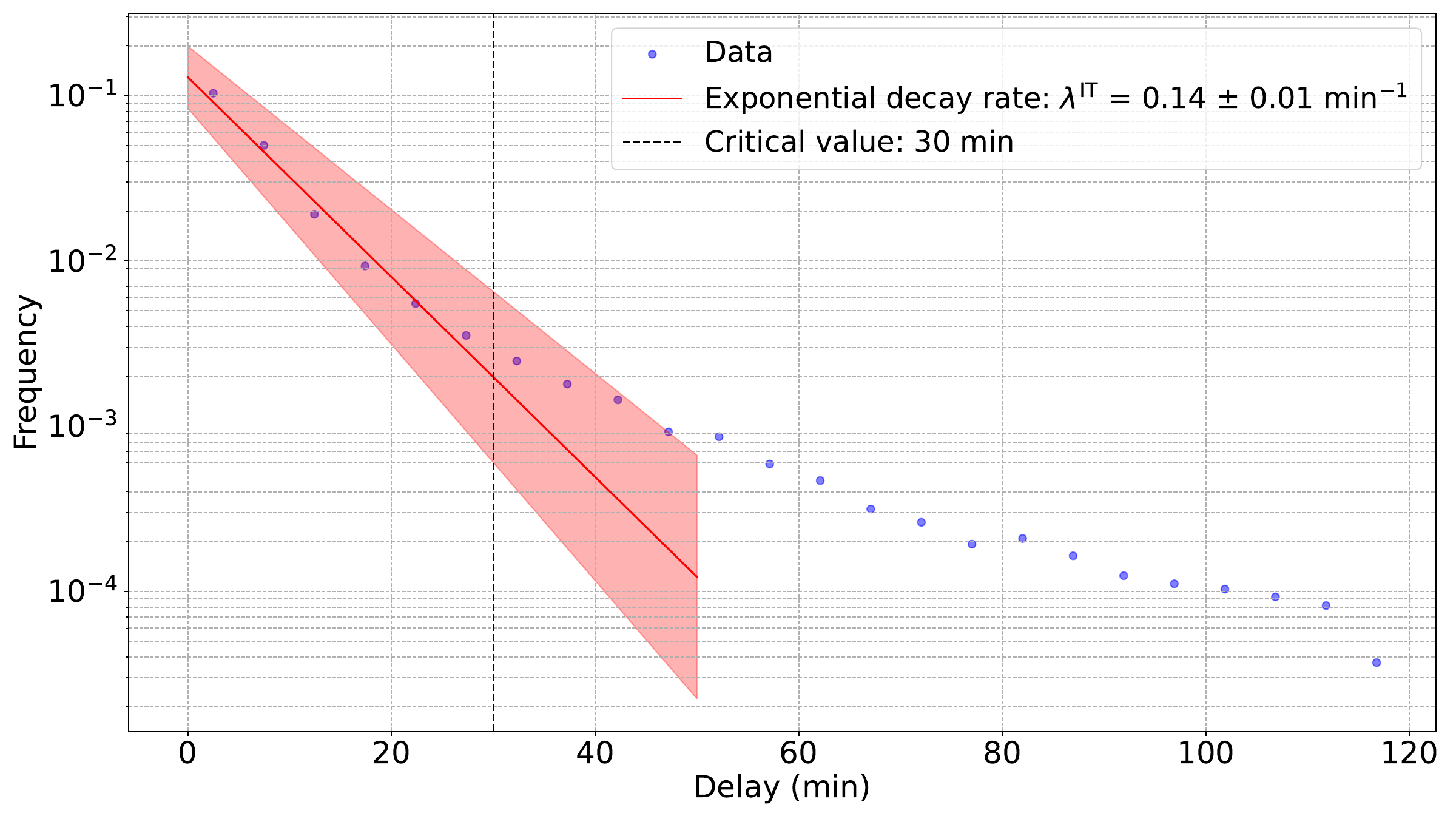}
    \caption{\textbf{Transition from exponential to power law decay of delay distribution for Italian high-speed trains.} The figure shows the distribution of train delays for Italian high-speed trains, which follows an exponential decay up to approximately 30 minutes and transitions to a power law decay at longer delays. The red line indicates the exponential fit, computed over the 0–30 minute interval and extended beyond this range with confidence level of $95\%$ to illustrate the deviation from exponential behavior at longer delays.}
    \label{fig:freccia-overall}
\end{figure}

\begin{figure}[tbp]
    \centering
    \includegraphics[width=\linewidth]{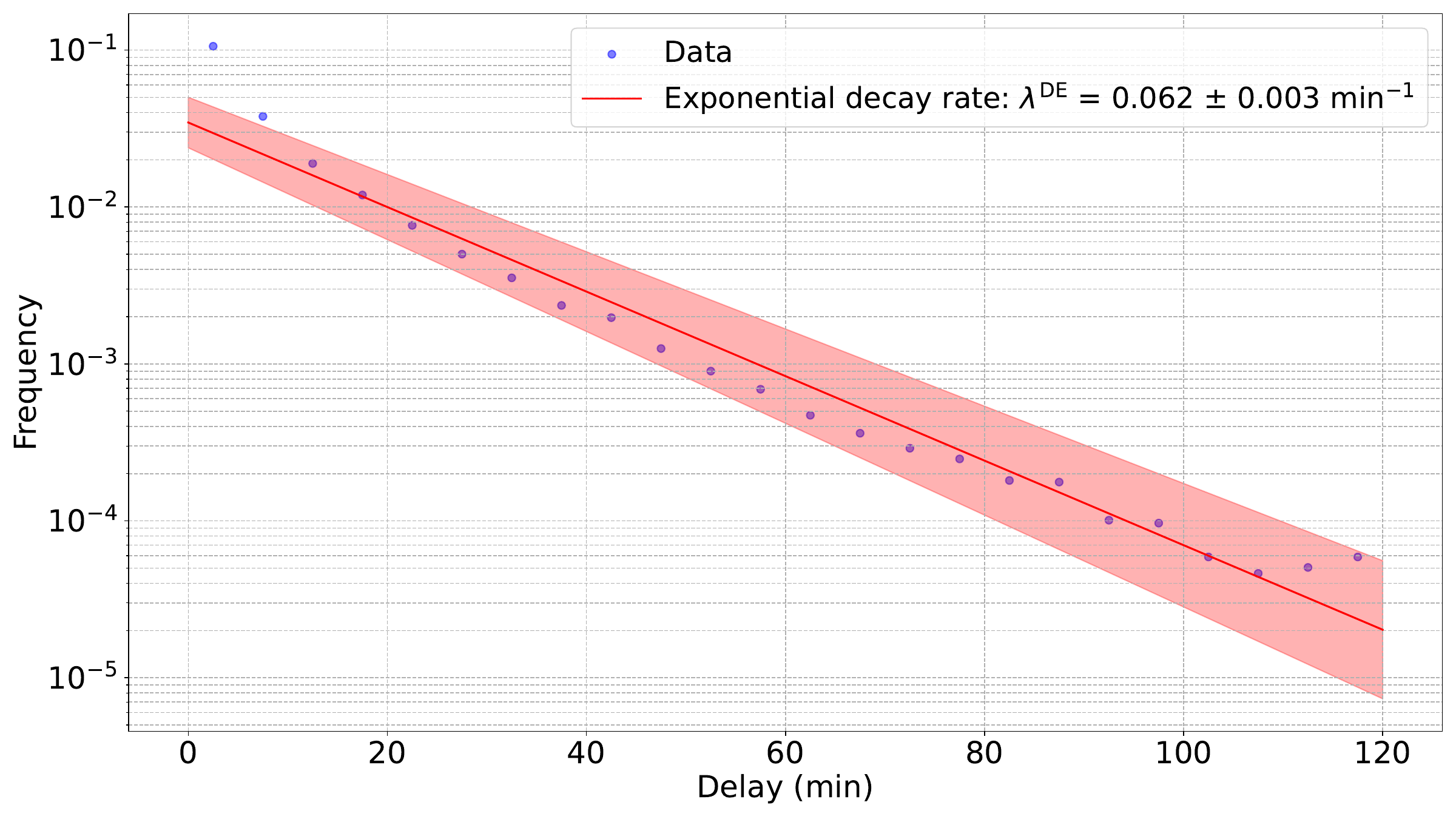}
    \caption{\textbf{Exponential decay of delay distribution for German high-speed trains.} The figure shows the distribution of train delays for German high-speed trains, which follows an exponential decay across the entire time interval. The red line represents the exponential fit, calculated and plotted over the same delay range with confidence level of $95\%$.}
    \label{fig:german_freccia}
\end{figure}

\begin{figure}[tbp]
    \centering
    \includegraphics[width=\linewidth]{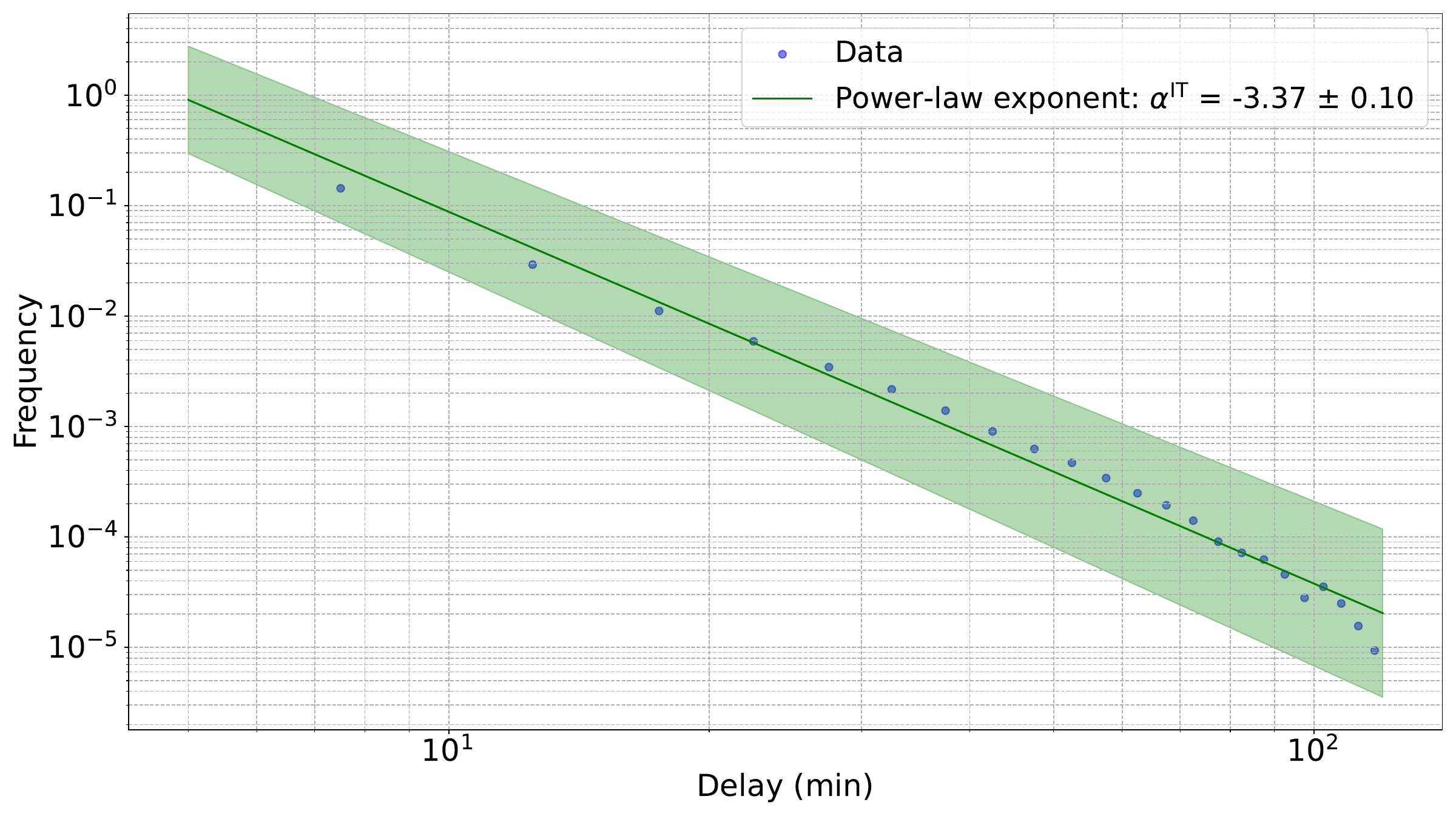}
    \caption{\textbf{Power law decay of delay distribution for Italian local trains.} The figure shows the distribution of train delays for Italian local trains, which follows an almost perfect power law decay. The green line represents the power law fit, calculated and plotted over the same delay range with confidence level of $95\%$.}
    \label{fig:regionale-overall}
\end{figure}

\begin{figure}[tbp]
    \centering
    \includegraphics[width=\linewidth]{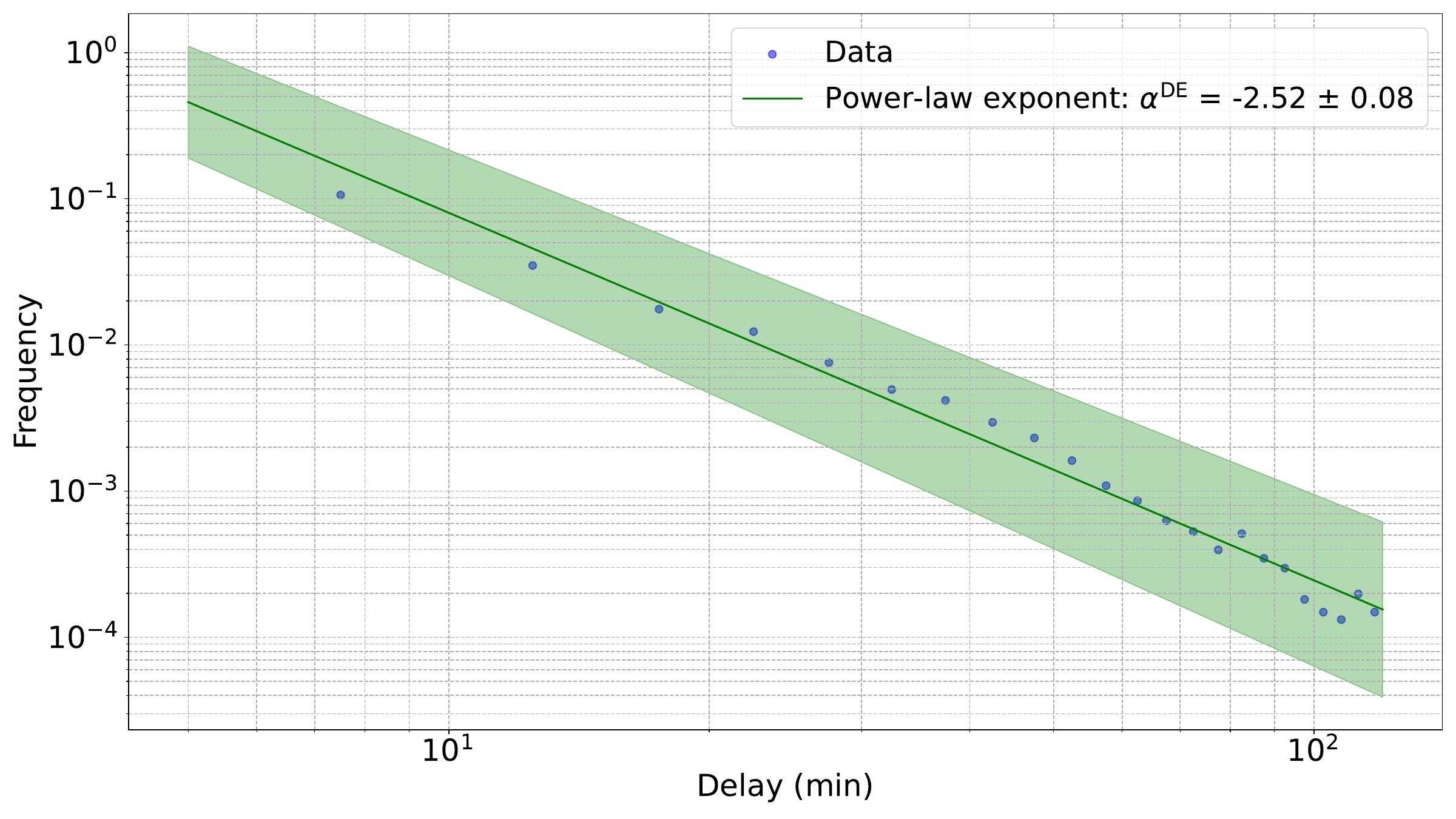}
    \caption{\textbf{Power law decay of delay distribution for German local trains.} The figure shows the distribution of train delays for German local trains, which follows an almost perfect power law decay. The green line represents the power law fit, calculated and plotted over the same delay range with confidence level of $95\%$.}
    \label{fig:german_regionale}
\end{figure}

We analyzed the real train delays of Italian and German railways during $458$ days. The high-speed train (high-priority agents) delays follow an exponential decay. Moreover for the Italian railway we notice a discrepancy in the exponential decay after $30$ minutes, which may indicate a switch to a power law distribution as shown in \autoref{fig:freccia-overall}. This is probably due to the drop of the priority enforced by refund policies in Italian high-speed trains. On the contrary, German trains keep exponential over the whole delay domain with a slower decay rate with respect to Italian ones: $\lambda^{\text{DE}} = \left(0.062 \pm 0.003\right)$ min$^{-1}$ vs $\lambda^{\text{IT}} = \left(0.14 \pm 0.01\right)$ min$^{-1}$ (cfr. \autoref{fig:german_freccia}).
Local train delays follow a power law distribution on both datasets: Italian ones have a greater exponents ($\alpha^{\text{IT}} = \left(-3.37 \pm 0.10\right)$ in \autoref{fig:regionale-overall}) with respect to the German ones ($\alpha^{\text{DE}} = \left(-2.52 \pm 0.08\right)$: cfr. \autoref{fig:german_regionale}).

We changed the model according to the connectivity of the realistic railway networks and we have introduced $6$ levels of priority according to the Italian trains classification. The train priorities are reported in \autoref{tabItCategories}. We performed simulations of the Italian train dynamics to check if the delay distributions of the high-speed and local trains follow the real ones.
The \emph{degree} of stations was evaluated from the scheduled trips of local trains and the \emph{management time} assuming the same value ($5$ minutes) for all the stations. 

\begin{figure}[tbp]
    \centering
    \includegraphics[width=\linewidth]{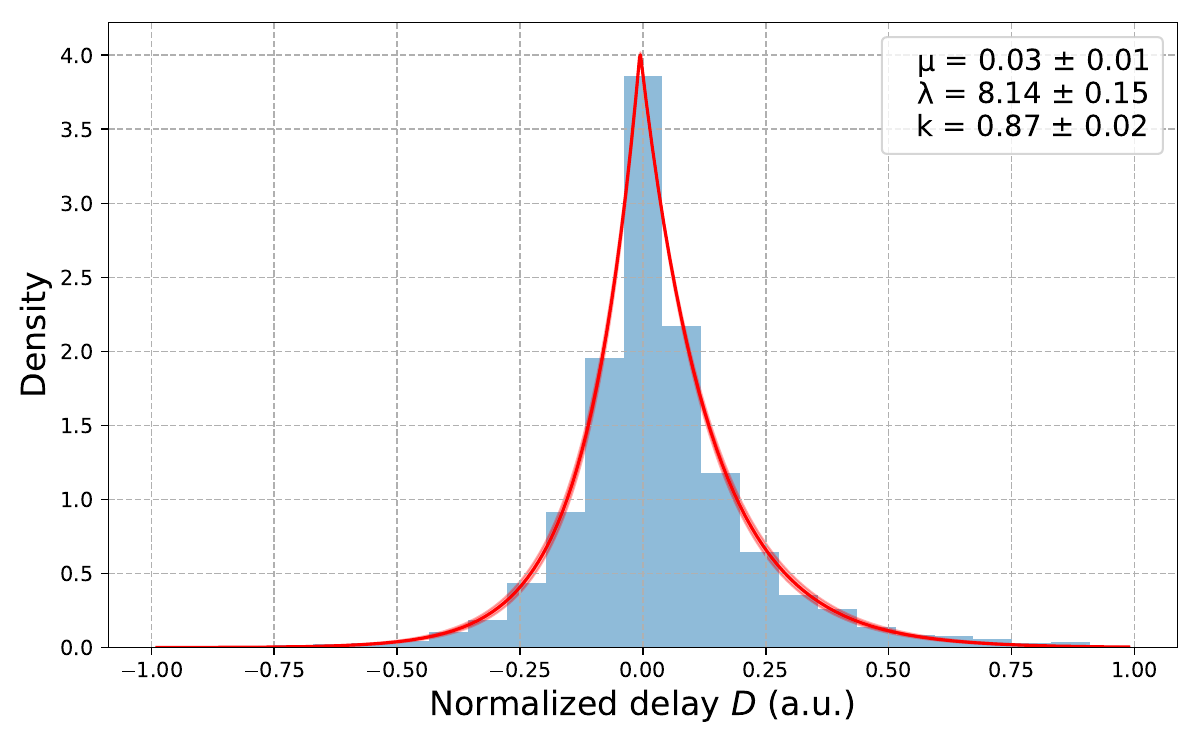}
    \caption{\textbf{Distribution of station-to-station delay ratios.} The figure shows the distribution of the ratio between the observed delay and the scheduled travel time for each station-to-station segment. The red line represents the fitted Asymmetric Laplace Distribution.}
    \label{fig:freccia-sts}
\end{figure}

\begin{figure}[tbp]
    \centering
    \includegraphics[width=\linewidth]{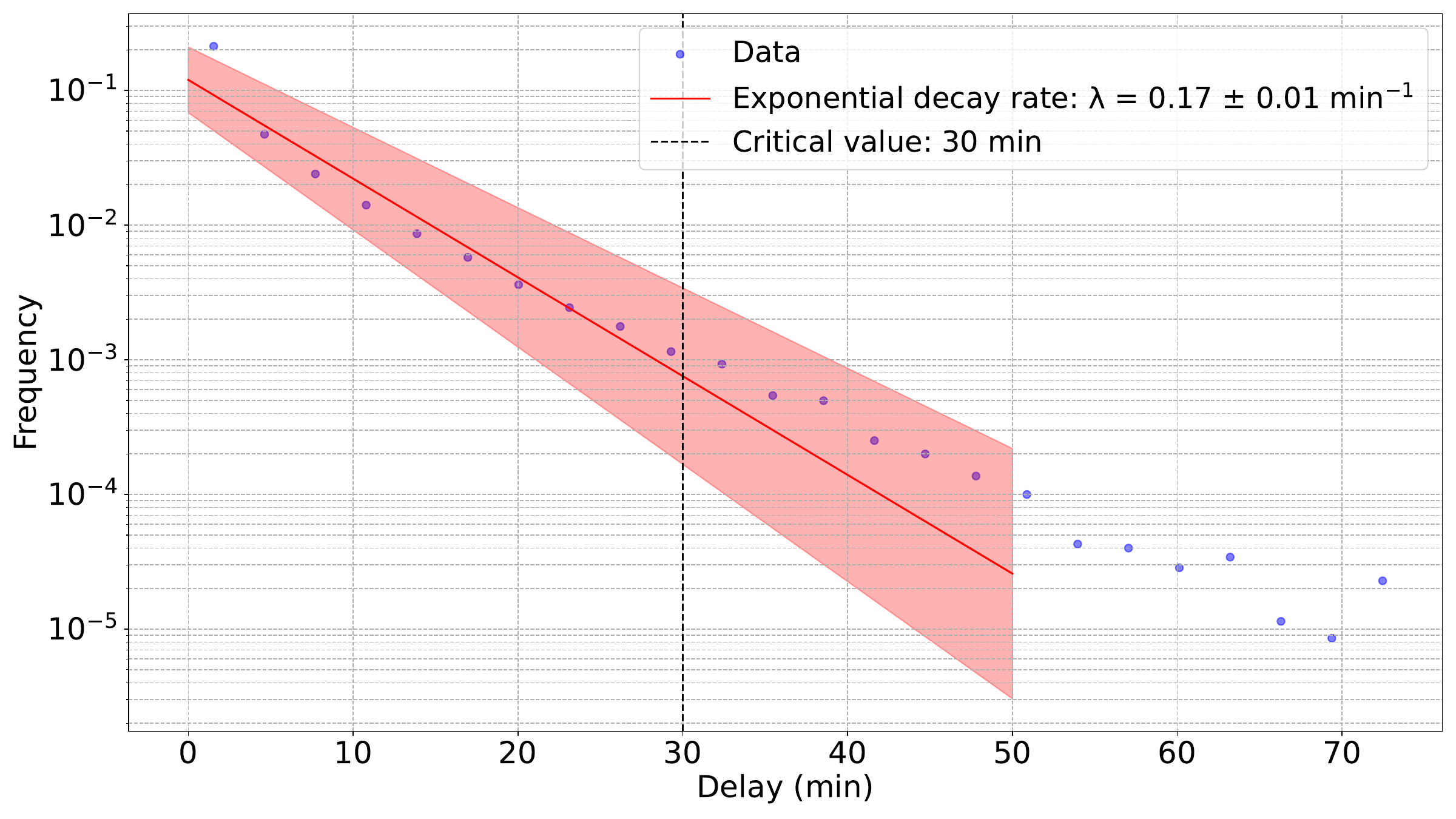}
    \caption{\textbf{Transition from exponential to power law decay in simulated Italian high-speed train delays.} The figure shows the distribution of simulated high-speed train delays, which follows an exponential decay up to approximately $30$ minutes and transitions to a power law decay at longer delays. The red line represents the exponential fit, calculated up to the critical value and plotted over a wider delay range with confidence level of $95\%$ to highlight the deviation from exponential behavior.}
    \label{fig:sfreccia}
\end{figure}

\begin{figure}[tbp]
    \centering
    \includegraphics[width=\linewidth]{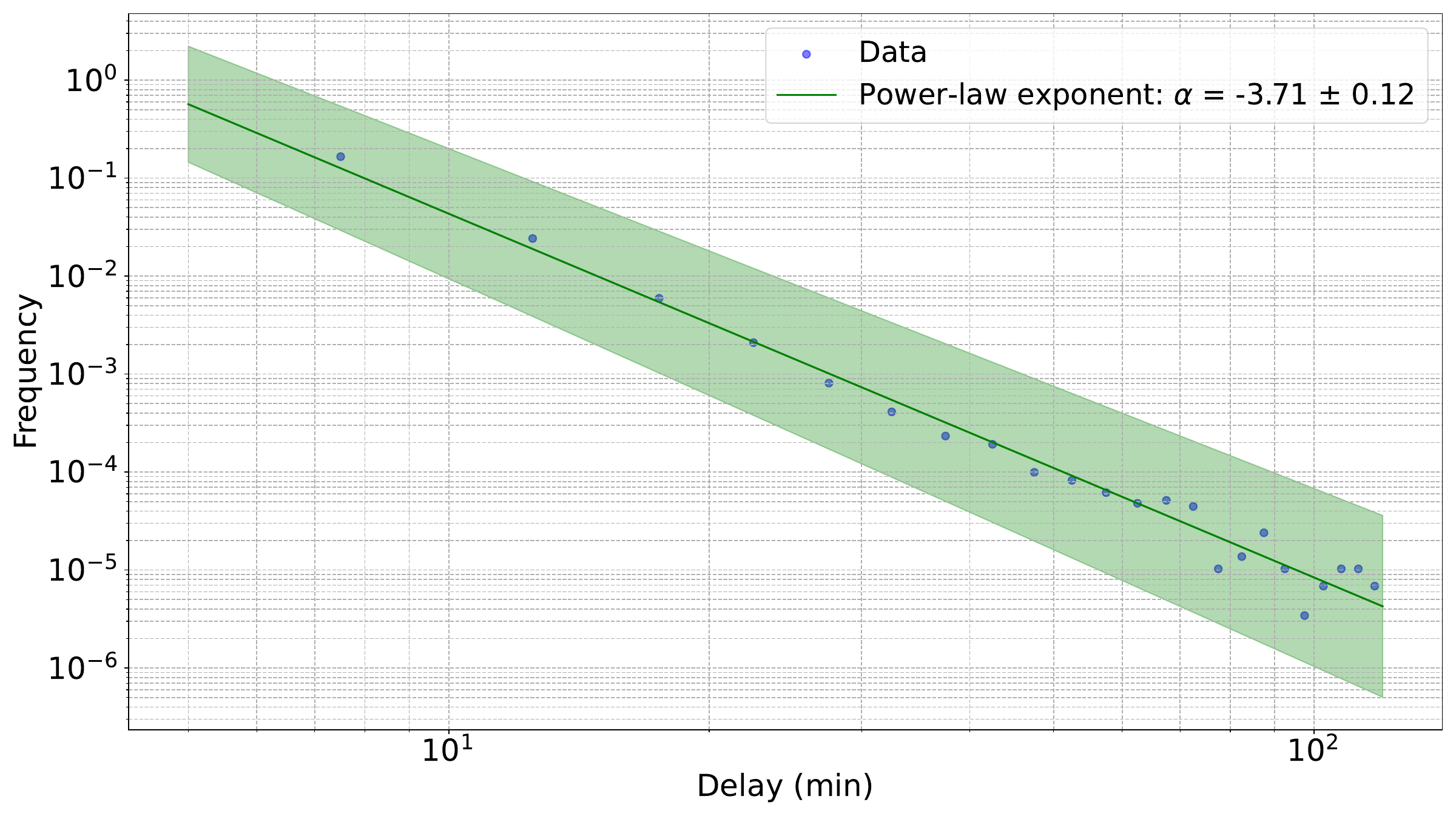}
    \caption{\textbf{Power law decay of delay distribution for simulated Italian local trains.} The figure shows the distribution of simulated train delays, which follows a power law decay. The green line represents the power law fit, calculated and plotted over the same delay range with confidence level of $95\%$.}
    \label{fig:sregionale}
\end{figure}

We added fluctuations using a Laplacian distribution to the travel time of the trains in order to reproduce the effective travel time.
The fluctuations follow the empirical distribution of delay and early of real trains (see \autoref{fig:freccia-sts}).
We performed $458$ independent simulations to compare the results with the available data on $458$ different days.
The simulation results reproduce the empirical delay distributions: the delay distribution for the high-speed trains has an exponential behavior with $\lambda = \left(0.17 \pm 0.01\right) \text{min}^{-1}$ on delays smaller than $30$ minutes and then it shows a power law tail after the priority drop (see \autoref{fig:sfreccia}).
Conversely, local train delays follow a power law distribution as shown in \autoref{fig:sregionale} with $\alpha = \left(-3.71 \pm 0.12\right)$.
The decaying rate and the exponent estimated by the simulation are consistent with the real data distributions.

\section*{Methods}\label{secMethods}
Our aim is to explain the emergence of heavy-tailed delay distributions in multi-layer networks governed by hierarchical priority rules. We propose a dynamical framework where agents move on a mobility network under finite nodal capacity constraints, where high-priority agents operate independently of low-priority traffic, while the latter suffers cumulative delays due to queuing interactions.

All data interpolations were performed using \emph{statsmodels package}~\cite{statsmodels} and consist of an ordinary least squares regression with a $95\%$ confidence interval.
The asymmetric Laplacian fit (see \autoref{fig:freccia-sts}) was performed using a non-linear least squares method implemented in the \emph{scipy package}~\cite{scipy}.

\subsection*{Model}\label{subsecMethodsModel}
The considered model is a multilayer network defined as $\mathcal{G} = \{V, E\}$, where $V$ denotes the set of nodes (transit hubs) and $E$ represents the edges (physical or logical links). The origin-destination dynamics on the network is performed by two distinct classes of agents: high-priority agents ($\mathcal{H}$) and low-priority agents ($\mathcal{L}$).

\subsubsection*{Nodal Dynamics and Capacity Constraints}
Each node $i \in V$ has a finite output transport flow governed by its degree $k_i$, which represents the number of parallel service channels. We introduce a characteristic management time $\tau$, defined as the minimum temporal interval required between successive agent departures from a single channel. The effective service rate $\mu_i$ of a node is expressed as: $\mu_i = \frac{k_i}{\tau}$.

The nodes are assumed to have infinite buffers for incoming agents and the departure process is regulated by a non-preemptive priority queue. Agents in class $\mathcal{H}$ are serviced according to their scheduled timestamps without interference from class $\mathcal{L}$. An agent $a \in \mathcal{L}$ is permitted to depart at its departure time $t$ only if a service channel is vacant and no agent $b \in \mathcal{H}$ is scheduled for departure within the temporal window $[t,\ t+\tau]$.

\subsubsection*{Travel Time Fluctuations}\label{subsecMethodsDelay}

Independently of network-induced congestion, each agent path undergoes an intrinsic travel time variability. For any link between nodes $i$ and $j$ with scheduled duration $T_{ij}$, the observed travel time is $T'_{ij} = T_{ij} \left(1 + D_{ij}\right)$, where $D_{ij}$ is a stochastic fluctuation.

We assume that the fluctuations are due to independent events so that the delays should following an exponential distribution: $\exp{\left(\lambda_{+}\right)}$. Similarly, early arrivals (negative delays) should follow another exponential distribution $\exp{\left(\lambda_{-}\right)}$ with a distinct rate parameter.
Since agents are statistically more likely to experience delays than advances, we also expect $\lambda_{+} < \lambda_{-}$.

The overall distribution of $D_{ij}$ is an \textit{Asymmetric Laplace Distribution}, whose probability density function (PDF) is given by:
\begin{equation}
\label{eqALpdf}
f(x;m,\lambda,k) = \frac{\lambda}{k+k^{-1}}e^{-\left(x-m\right)\lambda sk^s}
\end{equation}
where $s = sign\left(x-m\right)$.

\begin{proof}
    Starting from equation \eqref{eqALpdf} we can compute the characteristic function of the Asymmetric Laplace distribution using the definition~\cite{Jammalamadaka31122004}.
    For simplicity we put $m=0$, but the proof is indeed general.
    \begin{equation}
        \begin{split}
            \phi^{AL}\left(t;\lambda,k\right) &= \int_{-\infty}^{\infty}f(x;\lambda,k)e^{itx}dx = \\
            &= \frac{\lambda}{k + k^{-1}}\left(\int_{-\infty}^{0}e^{\left(it+\frac{\lambda}{k}\right)x}dx + \int_{0}^{\infty}e^{-\left(-it+\lambda k\right)x}dx\right) = \\
            &= \frac{\lambda}{k + k^{-1}} \left(\frac{1}{it+\frac{\lambda}{k}} + \frac{1}{-it+\lambda k}\right) = \frac{1}{1+\frac{itk}{\lambda}}\frac{1}{1-\frac{it}{\lambda k}}\ .
        \end{split}
        \label{eqALchf}
    \end{equation}
    The exponential characteristic function is indeed
    \begin{equation*}
        \phi^{EXP}\left(t;\lambda\right) = \frac{1}{1-\frac{it}{\lambda}}\ .
    \end{equation*}
    
    Given $X \sim \exp\left({\lambda_{+}}\right)$ and $Y \sim \exp\left({\lambda_{-}}\right)$, the characteristic function of $Z = X - Y$ is the product of the two characteristic function, namely
    \begin{equation}
        \phi\left(Z;\lambda_{+}, \lambda_{-}\right) = \frac{1}{1-\frac{iZ}{\lambda_{+}}}\frac{1}{1+\frac{iZ}{\lambda_{-}}}\ .
        \label{eqTOTchf}
    \end{equation}
    
    By comparing equations \eqref{eqALchf} and \eqref{eqTOTchf} one can find that $\lambda = \sqrt{\lambda_{+}\lambda_{-}}$ and $k = \sqrt{\frac{\lambda_{+}}{\lambda_{-}}}$.
\end{proof}

\subsubsection*{Toy model}
To study the impact of priority rules regardless of  specific topological features of the transport network, we use a simplified model defined on a $12 \times 12$ 2-layers synthetic grid.
In \autoref{fig:grid} we show the plot of the 2 layers grid structure. The implementation of the model is based on the following information.
\begin{itemize}
    \item \textbf{Grid Topology:} Nodes are classified by their degree: low-connected ($n=1\,012$, $k=2$), middle-connected ($n=385$, $k=4$), and high-connected ($n=144$, $k=8$). Class $\mathcal{H}$ agents are restricted to the high-connectivity subgraph, the fast layer; whereas class $\mathcal{L}$ agents traverse all available nodes (slow layer).
    \item \textbf{Simulation Parameters:} The system was initialized with $N_L = 18\,000$ low-priority agents and $N_H = 9\,000$ high-priority agents, with start times distributed across $1\,000$ arbitrary time units and distance-dependent node-by-node travel times, to make high-priority agents travel times higher with respect to the low-priority ones.
\end{itemize}

In \autoref{fig:sfr_gr_pr} is shown how high-priority agents delay distribution is weel interpolated by an exponential distribution, meanwhile low-priority agents delays, reported in \autoref{fig:srg_gr_pr}, follow a power law distribution.

To check the impact of the priority in the delay distributions, we ran again the previous simulation putting all agents in the same priority class, letting them move on the different network layers. In this way we still have $N_H$ agents on the fast layer and $N_L$ agents on the slow one, but without ordering preferences at junctions. The result, shown in \autoref{fig:sfr_gr_npr} and \autoref{fig:srg_gr_npr}, highlights how this time both distributions are exponential-like.

\subsection*{Application to real data}\label{subsecMethodsDataAnalysis}
We have generalized the toy model to simulate real complex railway networks where agents move with different priority rules at the main stations. The priority rules are implemented according to the available information.
In this way it is possible to perform a validation procedure using the model predictions on the delay distributions. 
We explicitly considered the railway networks in Italy and Germany and the corresponding priority rules.
In both cases, high-speed trains are high-priority agents that stop only on main stations, i.e. highly connected nodes, and move on the fast layer. Conversely,  local trains are low-priority agents that stop on every station: high-connected, middle-connected and low-connected nodes and they move on slow layer
We remark that real networks may have a multi-layer structure with more than two layers, as we can infer from the variety of train categories reported in \autoref{tabItCategories}: we expect the results to be coherent with the previous observations.

We first analyzed the empirical delay distributions using real data, checking whether our theoretical assumptions are consistent with the real observations.
Then, we performed simulations using real scheduled data on our model to test if it is able to reproduce the real dynamics.

We also take into account that, according to the Italia rules, if a high-priority agent delay is greater than a threshold value, its priority is downgraded with respect to other high-priority agents, thus remaining bigger than low-priority ones.
This fact is related to refund policies and we suppose it was introduced to avoid cascade effects.

\subsubsection*{Data}\label{subsecMethodsData}
We collected and analyzed empirical data from the Italian and German railway networks.
The Italian dataset comprises records from multiple dates in 2023 --specifically, September 19, 21, 26, 27, and 28, as well as October 6, 7, 9, 10, 11, and 13-- and a continuous period spanning from January 31, 2025, to April 23, 2026.
All Italian data were retrieved from the \textit{Viaggiatreno} portal (\url{https://www.viaggiatreno.it/}).
Complementary data for the German railway system, covering the months of February, March, and April 2025, were obtained from a publicly accessible online archive (\url{https://piebro.github.io/deutsche-bahn-statistics/}).

For the Italian railway network, the \textit{Viaggiatreno} platform provides detailed information essential to our study, including train category (which determines priority levels) and station-to-station delay times.
We remapped the original train categories into two aggregated groups: \emph{local} and \emph{high-speed}, as detailed in \autoref{tabItCategories}.
The priority levels reported are those directly used in our model and are not subject to further aggregation.
\emph{EuroCity} and \emph{Frecciarossa} trains are assigned the highest priority, while \emph{InterCity} services receive intermediate priority, and \emph{SFM} (metropolitan services) are designated the lowest priority.
For the German data set, trains were classified as \emph{high-speed} only if they belong to the \textit{ICE}, \textit{TGV}, or \textit{RJX} classes.
The impact of freight traffic was considered negligible for the purposes of this study.

The Italian railway network topology was reconstructed by analyzing train itineraries: $2\,073$ stations were identified as active, i.e. visited by at lest one train per observation day.

\subsubsection*{Delay distribution analyses}
Having the $x_{ij}$ delays in the Italian data, we analyzed them to asses their compatibility with the assumptions made in \emph{Travel Time Fluctuations}. An Asymmetric Laplace distribution fit is reported in \autoref{fig:freccia-sts} on travel time fluctuations in percentage. Notice that only small high-speed train delays are considered to minimize the influence of interactions with other trains or layers and show just the intrinsic travel fluctuations.
The resulting mean is close to zero $\mu = 0.03 \pm 0.01$, an asymmetry parameter $k = 0.87 \pm 0.02$ and $\lambda_{+} = 7.07 \pm 0.19$, $\lambda_{-} = 9.37 \pm 0.25$ confirms the $\lambda_{+} < \lambda_{-}$ hypothesis.

We then analyzed the distribution of the delays over entire trips and in both cases a power law distribution emerges for the delays of local trains
whereas an exponential distribution explains the delay of high-speed trains. But a different behavior emerges in the two considered cases. The distribution on 458 days of Italian data show that, for local trains, the delay distribution is well explained by a power-law, as we can see in \autoref{fig:regionale-overall}. For delays smaller than 30 minutes the high-speed train distributions is well described by an exponential law, but then one observes a fat tail (see \autoref{fig:freccia-overall}). This phenomenon is probably reflecting Trenitalia's operational policies, being 30 minutes the minimum required delay to qualify for a refund, according to the official guidelines (\url{https://www.trenitalia.com/it/informazioni/indennizzo-per-ritardodeltreno.html}). Conversely, the analyses over a three-month period of the German trains highlights a perfect exponential trend in the high-speed train case, \autoref{fig:german_freccia}, and a power-law behavior for local trains, \autoref{fig:german_regionale}. In this case, we do not see the priority drop, probably due to different regulations: German tickets are eligible for a refund only after 60 minutes (\url{https://int.bahn.de/en/booking-information/passenger-rights/legal-regulations}) and they have a general minimum cost requirement lower with respect to the Italian one.

\subsubsection*{Train Dynamics Simulations}
To test whether train delay distributions depends on their priorities, we simulated their trips using our model on a synthetic railway network in the case of Italian railway.
The Italian train routes are extracted from \textit{Viaggiatreno} data, saving stops and scheduled arrivals and departures.

We kept the same dynamics rules of the grid model, using the real railway network topologies and the real scheduled times for trains.
In particular, the networks are built starting from scheduled train routes: two stations are connected if they are consecutive stops of at least one train route.
Then, we set the number of trains that are able to depart simultaneously as the station degree.
This may produce a bias for which stations will have higher degree than the real number of tracks, but results will show this is not a limitation.

The management time of stations is tuned by taking into account the scheduled trains, allowing their departures without congestion if travel time fluctuations are suppressed.

High-speed trains lose priority after $30$ minutes in the simulations too, in order to reproduce the real results.

In \autoref{fig:sfreccia} and \autoref{fig:sregionale} the delay distributions of high-speed and local trains are showed, respectively.
Notice that high-speed train delays follow an exponential distribution until 30 minutes delay and then show a power law decay, while local train delays follow a power law across the entire plotted domain.

\section*{Discussion}\label{secConclusions}
In this work, we have shown that simple hierarchical rules, implemented via priority-based queue dynamics on a multi-layer network, can explain the heavy-tailed delay distributions observed in railway networks, without imposing any topological network constraint.
Indeed, the simulation results on a toy model with a two-layer grid network suggest that the priority rules are sufficient to change extreme events distributions, showing that the delay distribution of the low-priority agents is well interpolated by a power law distribution. This simple mechanism  suggests that power law distributions may reflect an universal properties of multi-layer transport network and that the exponent depends on the topological structure of the transport network, but further studies are necessary to prove this conjecture.

We have analyzed the empirical delay distribution of the Italian and German trains to highlight as an exponential distribution can explain the high-speed train delays whereas a power law distribution is observed for the local trains.

The main result of the paper is to show that adapting the model to simulate a realistic railway network we can quantitatively reproduce the empirical delay distributions for the Italian trains by modeling high-speed and local trains as agents with distinct priority levels and introducing empirically calibrated Laplacian-distributed fluctuations in inter-station travel times. Our simulations naturally yield exponential delay statistics for high-priority services and power law tails for lower-priority services. Although a complete validation of the model would require a study of its predictive capabilities,
these findings are consistent with the idea that simple rule sets, rather than elaborate interaction kernels or finely tuned network topologies, can generate complex macroscopic phenomena such as power law delays~\cite{gao2024}.

A direct comparison between the Italian and German datasets further highlights how operational policies imprint themselves on delay statistics.
The Italian network exhibits sharp cut-offs at $30$ minutes for high-speed trains: this threshold is directly tied to passenger refund policies.
On the other hand, the German system, lacks of analogous priority-change provisions due to different refund policies which are common for all train types and start at 60 minutes.
The result is an exponential or power law behavior across the entire delay domains.
This divergence is also reflected by a lower power law exponent for German local trains, indicating that aggressive prioritization of high-speed services can significantly exacerbate delays in local mobility.

We are aware that our model carries several simplifying assumptions that require future investigation. In particular for the toy model, level of congestion has not been analyzed neither studied to understand when a framework become sensitive to priority rules.
Further analyzes may be carried out to understand the statistics of very extreme events and how they affect the framework.

When we simulate realistic railway networks, our station-degree approximation abstracts away real track layouts and platform constraints; incorporating detailed infrastructure maps and direction-specific capacities could refine delay propagation patterns.
Moreover the instantaneous transit assumption at non-stop stations omits potential interactions at intermediate nodes; richer path information, including all intermediate halts, would enable more accurate modeling of congestion cascades.
We finally remark that freight traffic, which we treated as negligible, may play a significant role in mixed‐use corridors and should be explicitly accounted for.

In conclusion, by elucidating the mechanisms through which minimal priority‐based hierarchies engender emergent heavy-tails in delay statistics, this study contributes to a deeper understanding of complexity in engineered systems and offers practical insights into the design of operational policies for resilience enhancement and it offers actionable guidance for transport operators seeking to balance service quality with operational resilience.

\section*{Acknowledgements}

The authors thank Riccardo Barbieri for his technical support and expertize in automating the data extraction tools.

\section*{Author contributions statement}

Tommaso Rondini had the idea to analyze how the delays are distributed and how the network structure and policies influence them.

Tommaso Rondini and Gregorio Berselli performed the research.
The model was mainly developed by the former, while the data acquisition analyzes and mathematical formulation were mainly done by the latter.

Armando Bazzani and Mirko Degli Esposti provided many ideas, knowledge and solutions to the middle-step problems.

\section*{Data Availability Statement}
The datasets generated and/or analyzed during the current study are not publicly available due to copyright and third-party website terms of use, but are available from the corresponding author on reasonable request.

\section*{Funding Statement}
This research received no external funding.

\clearpage

\begin{table}[tbp]
    \centering
    \begin{tabular}{@{}lcc@{}}
        \toprule
        \textbf{ViaggiaTreno Category} & \textbf{Category} & \textbf{Priority Level} \\
        \midrule
        EC             & \emph{High Speed} & 6 \\
        Frecciarossa    & \emph{High Speed} & 6 \\
        Frecciaargento  & \emph{High Speed} & 5 \\
        Frecciabianca   & \emph{High Speed} & 5 \\
        IC             & \emph{Local} & 3 \\
        ICN            & \emph{Local} & 3 \\ 
        RV             & \emph{Local} & 2 \\
        Regionale      & \emph{Local} & 1\\
        SFM            & \emph{Local} & 0 \\
        \bottomrule
    \end{tabular}
    \caption{\textit{Viaggiatreno} train categories mapped into \emph{high-speed} and \emph{local} with an integer priority level.}
    \label{tabItCategories}
\end{table}

\end{document}